\documentclass[twocolumn]{article}
\advance\topmargin-1in\advance\oddsidemargin-1in
\evensidemargin\oddsidemargin

\makeatletter
\def\@normalsize{\@setsize\normalsize{12pt}\xpt\@xpt
\abovedisplayskip 10pt plus2pt minus5pt\belowdisplayskip \abovedisplayskip
\abovedisplayshortskip \z@ plus3pt\belowdisplayshortskip 6pt plus3pt
minus3pt\let\@listi\@listI}

\def\section{\@startsection {section}{1}{\z@}{20pt plus 2pt minus 2pt}
{8pt plus 2pt minus 2pt}{\centering\normalsize\sc
\edef\@svsec{\thesection.\ }}}
\def\thesection{\Roman{section}}

\def\subsection{\@startsection {subsection}{2}{\z@}{16pt plus 2pt minus 2pt}
{6pt plus 2pt minus 2pt}{\normalsize\sl
\edef\@svsec{\thesubsection.\ }}}
\def\thesubsection{\Alph{subsection}}

\long\def\@makecaption#1#2{
\vskip10pt\begin{center} #1 #2 \end{center}\par\vskip 1pt}
\def\fnum@figure{\raggedright{\footnotesize Fig. \thefigure }.%
\footnotesize}
\def\fnum@table{\footnotesize TABLE \thetable\\\footnotesize\sc}
\def\thetable{\Roman{table}}

\makeatother

\usepackage{graphicx}
\graphicspath{{./}{./figs/}}
\usepackage{algorithm}
\usepackage{algorithmic}

\usepackage{amsthm}

\newcommand{\tabincell}[2]{\begin{tabular}{@{}#1@{}}#2\end{tabular}}

\begin{document}
\date{}

\title{\Large\textbf{Multi-Voltage and Level-Shifter Assignment Driven Floorplanning}}


\author{Bei Yu\thanks{Bei Yu is with the Department of Computer Science and Technology, Tsinghua University, Beijing, China~~100084 (e-mail: disyulei@gmail.com)}
 \ , \ \ \ \ \ \  Sheqin Dong\thanks{Sheqin Dong is with the Department of Computer Science and Technology, Tsinghua University, Beijing, China~~100084 (e-mail: dongsq@mail.tsinghua.edu.cn)}
 \ \ \ \ and  \ \ Satoshi GOTO\thanks{Satoshi GOTO is with the Graduate School of IPS, Waseda University, Kitakyushu, Japan~~808-0135 }\\
} \maketitle \thispagestyle{empty}

{\small\textbf{Abstract}---
 As technology scales, low power design has become a significant
requirement for SOC designers. Among the existing techniques,
Multiple-Supply Voltage (MSV) is a popular and effective method to
reduce both dynamic and static power. Besides, level shifters
consume area and delay, and should be considered during
floorplanning. In this paper, we present a new floorplanning system,
called MVLSAF, to solve multi-voltage and level shifter assignment
problem. We use a convex cost network flow algorithm to assign
arbitrary number of legal working voltages and a minimum cost flow
algorithm to handle level-shifter assignment. The experimental
results show MVLSAF is effective.}

{\small\textbf{Index Terms---Voltage-Island, Multi-Voltage
Assignment, Level Shifter Assignment, Floorplanning}

\section{\textbf{INTRODUCTION}}

As technology scales, low power design has become a significant
requirement for system-on-chip designers. Many techniques were
introduced to deal with power optimization. Among the existing
techniques, Multiple-Supply Voltage (MSV) is one of the most
effective methods for both dynamic and static power reduction while
maintaining performance. In the MSV design, one of the most
important problem is voltage assignment: timing critical modules are
assigned to higher voltage while noncritical modules are assigned to
lower voltage, so the power can be saved without degrading the
overall circuit performance.

There are a number of previous works addressing voltage assignment
in floorplanning. Among these works, voltage assignment is
considered at various stages, including
pre-floorplanning\cite{ICCD05,ICCAD06LEE}; during
floorplanning\cite{ICCAD07MA,DAC08,ICCAD08MA,GLSVLSI09YU}; and
post-floorplaning\cite{ASPDAC07,ICCAD07LEE,ISPD09LEE}.

\textit{Level-shifter} \cite{ICCAD02} has to be inserted to an
interconnect when a lower voltage module drives a higher voltage
module or a circuit may suffer from excessive short-circuit current
energy. From \cite{ICCAD06LEE,GLSVLSI09YU} we can observe that the
number and the area of level shifters can not be ignored when
modules increase. As a result, level-shifters may cause performance
and area overhead, and should be considered during floorplanning.
Accordingly, MSV aware floorplanning includes two major problems:
voltage assignment and level shifter assignment, which make the
design process much more complicated.

Lee et al.\cite{ICCAD06LEE} handle voltage assignment by dynamic
programming, and level shifters are inserted as soft blocks. An
approach based on ILP is used in \cite{ICCAD07LEE} for voltage
assignment at the post-floorplanning stage. To make use of physical
information among modules during floorplanning, Ma et
al.\cite{ICCAD08MA} transform voltage assignment problem into a
convex cost network flow problem. However, their approach consider
neither level-shifters' area overhead nor level-shifters' physical
infomation.

Yu et al.\cite{GLSVLSI09YU} use a convex cost network flow algorithm
to assign voltage and a minimum cost flow algorithm to handle
level-shifter assignment which considers level-shifters' positions
and areas. However, their work can only assign two legal working
voltages. Besides, level shifters are assumed to be soft modules and
ratios can not controlled well.

In this paper, we propose a new floorplanning system MVLSAF, which
is extended from \cite{GLSVLSI09YU}. At floorplanning phase, we use:
a convex cost network flow algorithm to assign multi-voltages; a
minimum cost flow algorithm with more accurate model to assign level
shifters.

The remainder of this paper is organized as follows. Section 2
defines the voltage-island driven floorplanning problem. Section 3
presents our algorithm flow. Section 4 reports our experimental
results. At last, Section 5 concludes this paper.

\begin{figure*}[tbh]
\begin{center}
\includegraphics[width=0.75\textwidth]{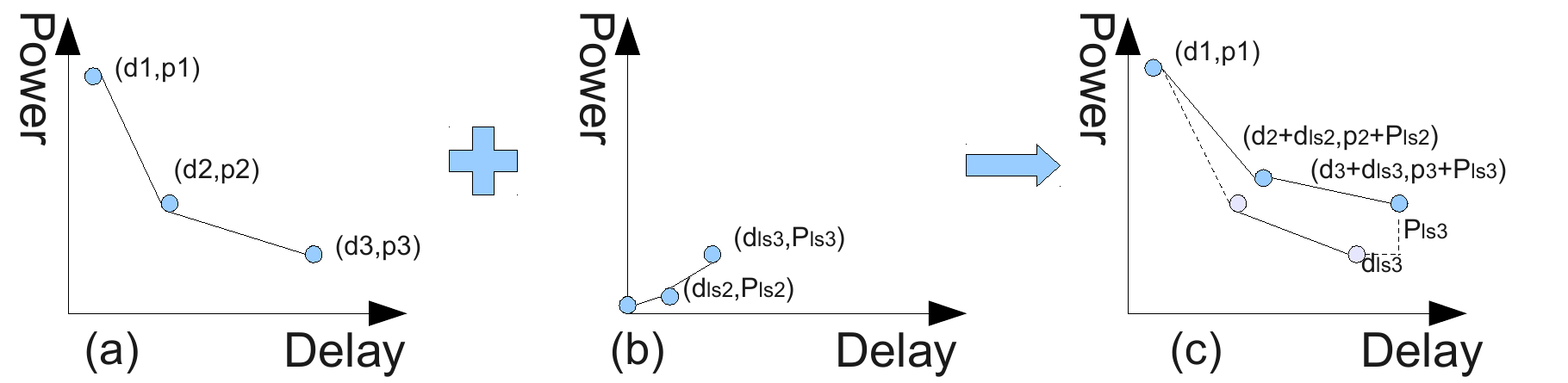}.
 \caption{Delay Power Curve of (a)module with three legal voltages; (b)corresponding level shifter. (c)modified DP-Curve of module.}
\label{e1}
\end{center}
\end{figure*}

\section{\textbf{PROBLEM FORMULATION}}
\newtheorem{define}{Definition}
\newtheorem{problem}{Problem}


\begin{define}[DP-Curve]The power-delay tradeoff of each module is
represented by a DP-Curve $\{(d_1,p_1), (d_2,p_2), \dots, (d_k,
p_k)\}$, where each pair $(d_i, p_i)$ is the corresponding delay and
power consumption when module is operated at voltage
$v_i$(Fig.\ref{e1}(a)).
\end{define}

We assume that power is a convex function of delay when each point
$(d_i, p_i)$ is connected to its neighboring point(s) by a linear
segment in the DP Curve. Besides, each level shifter has its own
DP-Curve ($(d_i, p_i)$ is delay and power consumption when it is
driving voltage $i$). Lower voltage module needs bigger level
shifter to drive other modules. Since bigger level shifter consumes
more delay and power, we assume the level shifter's DP-Curve is also
convex(as shown if Fig.\ref{e1}(b)). We extend the problem in
\cite{GLSVLSI09YU} to Multi-Voltage and Level-Shifter Assignment
driven Floorplanning (MVLSAF):

\begin{problem}{(MVLSAF)}
We are given following input to generate floorplanning result:
minimize the area, power cost and wirelength; satisfying timing
constraint; insert all the level-shifters in need.
\begin{flushleft}
\begin{tabular}{p{0.06cm}ll}
 1)&A set of modules, each module has its DP--curve.\\
 2)&A netlist $\hat G = (\hat V, \hat E)$ and timing constraint $T_{cycle}$.\\
 3)&Level-shifter's area, ratio and DP-Curve.\\
 4)&Number of legal working voltage $k$.
\end{tabular}
\end{flushleft}
\end{problem}

\section{\textbf{ALGORITHM of MVLSAF}}
Our work is similar to that presented in \cite{GLSVLSI09YU}, in
which a two phases framework is presented to deal with voltages and
level-shifters assignment during floorplanning. However, our
approach differs from \cite{GLSVLSI09YU} in the following ways:
during voltage assignment, we support arbitrary number of legal
voltages allowing further power reductions in certain applications;
during level-shifter assignment, we adopt more accurate model to
calculate possible level-shifter number in white space, which
control level shifter under certain H/W ratio.

\subsection{Multi-Voltages Assignment}
\newtheorem{lemma}{LEMMA}
\newtheorem{Theorem}{THEOREM}
To take consumption of level-shifter into consider, for each module,
we modify its DP-Curve: replace each pair $(d_i, p_i)$ by
$(d_i+d_{ls}^i, p_i+p_{ls}^i)$, where $(d_{ls}^i, p_{ls}^i)$ islevel
shifter's delay and power consumption. Modified DP-Curve is shown in
Fig.\ref{e1}(c).

\begin{lemma}
$f(x)$ is convex $\iff f(x_1+x_2)<\frac{f(x_1)+f(x_2)}{2}$, $\forall
x_1, x_2 \in Z$.
\end{lemma}
\begin{lemma}
If $f(x)$ and $g(x)$ are convex, then $P(x)=f(x)+g(x)$ is also
convex.
\end{lemma}

\begin{Theorem}
Modified DP-Curve is piecewise linear convex function with integer
breakpoints, and we can apply convex cost flow algorithm to solve
voltage assignment problem\cite{FLOW03}.
\end{Theorem}

\begin{figure}[tb]
\centering
\includegraphics[width=0.4\textwidth]{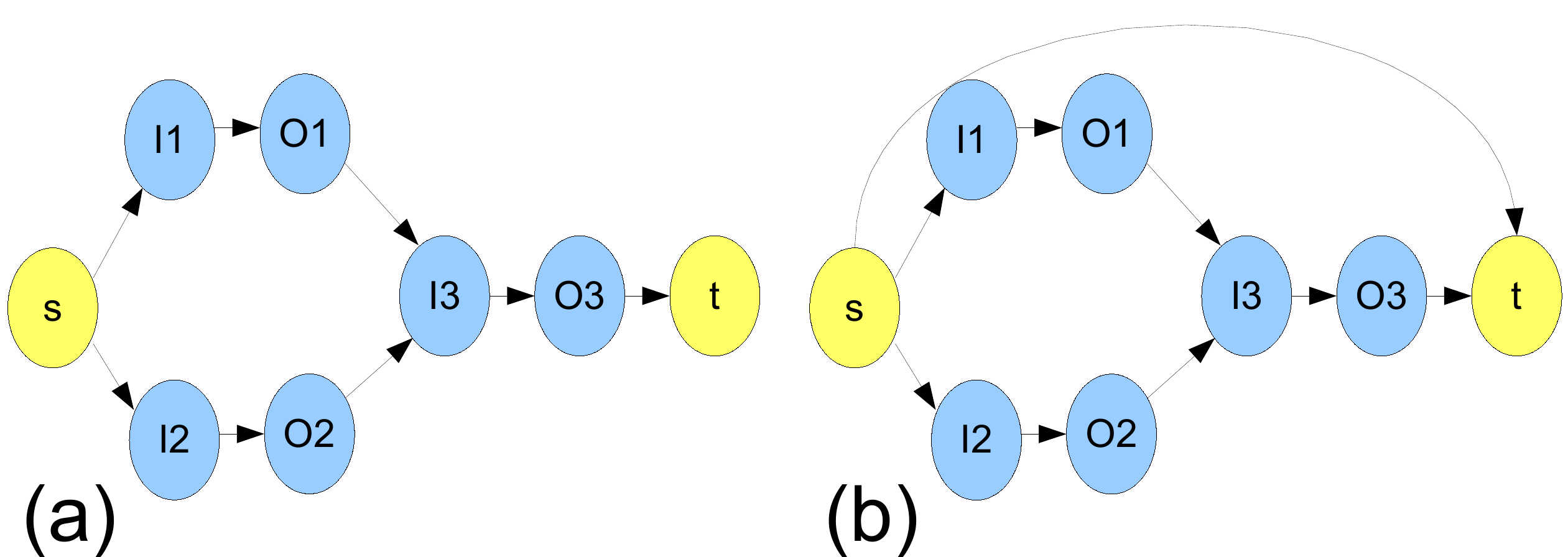}
\caption{~(a)$\bar G=\{\bar V,\bar E\}$, after adding nodes $s, t$
and diving nodes $N_i$ into $I_i$ and $O_i$ (b)Transformed $\bar
G=\{V, E\}$ by adding edge $e(\bar s,\bar t)$ to remove constraint
$\mu_t-\mu_s \leq T_{cycle}$ in equation (\ref{eq:1}).}\label{DAGs}
\end{figure}

Given netlist $\hat G$, we translate it into $\bar G =(\bar V, \bar
E)$(adding start node $s$ and end node $t$, dividing each node $n_i
\in \hat G$ into 2 nodes $I_i$ and $O_i$, as shown in
Fig.\ref{DAGs}(a)). So $\bar V = \{s, t, I_1, O_1,  I_2, O_2, \dots,
I_m, O_m \}$. And $I_i$ is connected to $O_i$ by a directed edge. We
denote these new created edges $\{e(I_i,O_i)|I_i, O_i\in \bar V\}$
as $\bar E_1$, denote edges $\{e(s,I_k)|I_k\in \bar V\}$ as $\bar
E_3$, and other edges as $\bar E_2$, and $\bar E=\bar E_1\cup \bar
E_2\cup \bar E_3$.

The mathematical program of voltage assignment is in
(\ref{eq:1}),where $d_{ij}$ is delay from node i to node j.
\begin{equation}\label{eq:1}
    Minimize \sum_{e(i,j)\in{\bar{E}}}P_{ij}(d_{ij})
\end{equation}
\begin{displaymath}
    s.t. \left\{
      \begin{array}{lll}
          \mu_j-\mu_i \geq d_{ij}         & \forall e(i,j)\in \bar{E}   & (1a)\\
          \mu_t-\mu_s \leq T_{cycle}      &                             & (1b)\\
          d_{ij}\in \{d_{ij}^1,d_{ij}^2, \dots, d_{ij}^k\} &\forall e(i,j)\in \bar E_1   & (1c)\\
          d_{ij}=delay_{ij}               & \forall e(i,j) \in \bar E_2 & (1d)\\
          d_{ij}=0                        & \forall e(i,j) \in \bar E_3 & (1e)\\
      \end{array}
    \right.
\end{displaymath}

We can incorporate constraints $(1b)$ and $(1a)$ by transforming
$(1b)$ into $\mu_s-\mu_t \geq -T_{cycle}$, and define $d_{st}$, s.t.
$\mu_t-\mu_s = d_{st}\quad \&\quad d_{st}\leq T_{cycle}$.
Accordingly, $\bar E_3=\{\bar E_3 \cup e(s,t)\}$, and the
transformed DAG $\bar G$ is shown in Fig.\ref{DAGs}(b). Besides, we
dualize the constraints $(1a)$ using a nonnegative Lagrangian
multiplier vector $\bar x$, obtaining the following Lagrangian
subproblem:
\begin{equation}
  L(\vec{x})=\textrm{min} \sum_{e(i,j)\in \bar E}[P_{ij}(d_{ij})+x_{ij}d_{ij}]
\end{equation}

We define function $H_{ij}(x_{ij})$ for each $e(i,j)\in E$ as
follows:
\begin{equation}
  H_{ij}(x_{ij})=\textrm{min}_{d_{ij}}\{P_{ij}(d_{ij})+x_{ij}d_{ij}\}
\end{equation}

\begin{Theorem}
The function $H_{ij}(x_{ij})$ is a piecewise linear concave function
of $x_{ij}$, and $\forall e(i, j)\in E_1$, $H_{ij}(x_{ij})$ is
described in the following manner: {\setlength\arraycolsep{2pt}
\begin{eqnarray}
H_{ij}(x_{ij}) & =& \left \{
                 \begin{array}{ll}
                   P_{ij}(d_{ij}^k)+d_{ij}^kx_{ij} & 0 \leq x_{ij}\leq b_{ij}(k)\\
                   \dots \\
                   P_{ij}(d_{ij}^q)+d_{ij}^qx_{ij} & 0 \leq x_{ij}\leq b_{ij}(q)\\
                   \dots \\
                   P_{ij}(d_{ij}^1)+d_{ij}^1x_{ij} & k \leq x_{ij}\\
                 \end{array}
               \right.
\end{eqnarray}}
where
$b_{ij}(q)=\frac{P_{ij}(d_{ij}^{q-1})-P_{ij}(d_{ij}^q)}{d_{ij}^q-d_{ij}^{q-1}}$.
\end{Theorem}


To transform the problem into a minimum cost flow problem, we
construct an expanded network $G' = (V', E')$. There are three kinds
of edges to consider:
\begin{flushleft}
\begin{itemize}
\item
  $e(i,j)$ in E1:we introduce $k$ edges in $G'$, and the costs of these
  edges are: $-d_{ij}^k, -d_{ij}^{k-1}, \dots -d_{ij}^1$; upper capacities:
  $b_{ij}(k), b_{ij}(k-1)-b_{ij}(k), b_{ij}(k-2)-b_{ij}(k-1), \dots
  M-b_{ij}(2)$, where $M$ is a huge confficient; lower capacities are both 0.
\item
  $e(i,j)$ in E2: cost, lower and upper capacity is $-d_{ij}$, 0, M.
\item
  Edge in E3: two edges are introduced in $G'$, one with cost, lower
  and upper capacity as ($-K_j, -M, 0$), another is ($0,0,M$).
\end{itemize}
\end{flushleft}

Using the cost-scaling algorithm, we can solve the minimum cost flow
problem in $G'$. For the residual network $G(x^*)$ and solve a
shortest path problem to determine shortest path distance $d(i)$
from node $s$ to every other node. By implying that $\mu(i) = d(i)$
and $d_{ij} = \mu(i) - \mu(j)$ for each $e(i,j) \in E_1$, we can
finally solve voltage assignment problem.

\subsection{Level-shifters Assignment}
\begin{table}[bt]\label{table:notation}
\centering \caption{Notation used in LS Assignment}
\begin{tabular}{|l|l|}
 \hline
 \hline $r_j$    & Room containing module $j$\\
 \hline $R$ & Set of rooms, $R = \{ r_1, r_2, \dots, r_m\}$\\
 \hline $LS$ & Set of LSs, $LS = \{ ls_1, ls_2, \dots, ls_{n}\}$\\
 \hline $p_{jk},k=1,2,3$ & Three parts of white spaces in $r_j$.\\
 \hline \hline
\end{tabular}
\end{table}
\begin{figure}[tb]
\centering
\includegraphics[width=0.5\textwidth]{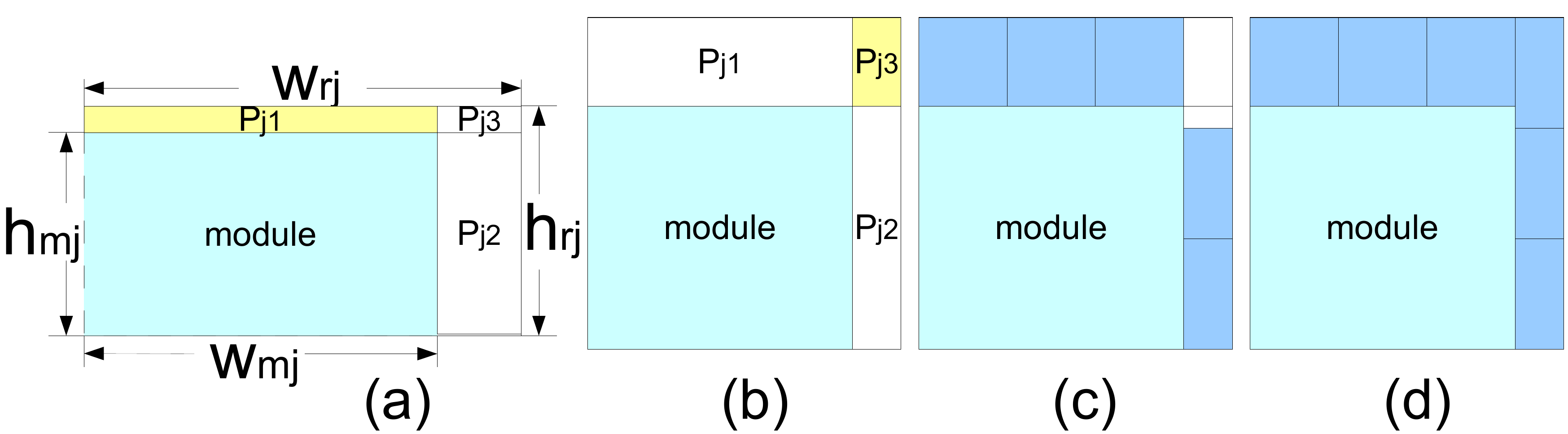}
\caption{~(a)Area of $p_{j1}$ is bigger than area of level shifter
but is too narrow to insert.~(b)$p_{j3}$ can merge with either
$p_{j1}$ or $p_{j2}$.~(c)If $p_{j3}$ merges with $p_{j1}$, can
insert only 5 level shifters.~(d)If $p_{j3}$ merges with $p_{j2}$,
can insert totally 6 level shifters.}\label{WhiteSpace}
\end{figure}

After voltage assignment, each module is assigned a voltage, then
the number of level shifters $n$ is determined. In level shifter
assignment we carry out minimum cost flow algorithm to try to assign
every level-shifters one room. Compared with \cite{GLSVLSI09YU}, we
present a more accurate model estimating possible number of level
shifters in white space to avoid too much ratio.

We construct a network graph $G^* = (V^*, E^*)$, and then use a
min-cost max-flow algorithm to determine which room each level
shifter belong to.

\begin{flushleft}
\begin{itemize}
\item
  $V^* = \{s, t\}\cup LS \cup R$.
\item
  $E^* = \{ (s, ls_i)|ls_i\in LS\} \cup \{ ( ls_i, r_j) | \forall fr_{ij}=1 \} \cup \{( r_j, t)| r_j \in R \}$.
\item
  Capacities: $C(s,ls_i) = C(ls_i, r_j)=1, C(r_j,
  t ) = NumLS(j)$.
\item
  Cost: $F(s, ls_i)=0, F(r_j, t)=0; F(ls_i, r_j)=F_{ij}$.
\end{itemize}
\end{flushleft}
where $fr_{ij} = \left \{
    \begin{array}{ll}
      1, & \textrm{if $ls_i$ can be inserted into $r_j$}\\
      0, & \textrm{others}\\
    \end{array}
  \right.$ and $F_{ij}$ can refer to \cite{GLSVLSI09YU}.

The algorithm of $NumLS(i)$, which accurately estimates number of
possible level shifters in white space, is shown in Algorithm
\ref{alg:num}. In room $r_j$, white space is split into at most
three parts: $p_{j1}, p_{j2}, p_{j3}$(as shown in
Fig.\ref{WhiteSpace}).

\begin{algorithm}[htb]
 \caption{NumLS(i)}
 \label{alg:num}
\begin{algorithmic}[1]
 \STATE Initialize $p_{i1}, p_{i2}, p_{i3}$;
 \STATE $a_{ls} \leftarrow $ area of level shifter;
 \IF{$p_{i1}$ is too narrow}
 \STATE $areaof(p_{i1}) \leftarrow 0$;
 \ENDIF
 \IF{$p_{i2}$ is too narrow}
 \STATE $areaof(p_{i2}) \leftarrow 0$;
 \ENDIF
 \IF{$ areaof(p_{i1})\  \% \ a_{ls} > areaof(p_{i2})\  \% \ a_{ls}$}
   \STATE $p_{i1} \leftarrow p_{i1} \cup p_{i3}$;
 \ELSE
   \STATE $p_{i2} \leftarrow p_{i2} \cup p_{i3}$;
 \ENDIF
 \RETURN $\lfloor \frac{areaof(p_{i1})}{a_{ls}} \rfloor + \lfloor \frac{areaof(p_{i2})}{a_{ls}} \rfloor$;
\end{algorithmic}
\end{algorithm}

It can be shown that any flow in the network $G^*$ assigns level
shifters to white spaces. The minimum cost flow algorithm can be run
in polynomial time\cite{FLOW03}.

After level-shifter assignment, level shifters that can not be
assigned to any room are belong to set $ELS$. We use heuristic
method to assign level shifters in $ELS$, so more level shifters in
$ELS$, more ILO(Interconnect Length Overhead\cite{GLSVLSI09YU}). If
white space $ws_i$ is too narrow, then level shifters being assigned
to $ws_i$ are all belong to $ELS$. More accurate model we used
(Algorithm 1) can reduce the number of level shifters in $ELS$ and
then reduces ILO.

\section{\textbf{EXPERIMENTAL RESULTS}}
\begin{table*}[tb]\label{table:result}
\centering \caption{The Comparison Between the VLSAF and the
Previous Work \cite{GLSVLSI09YU} }
 \footnotesize
\begin{tabular}{|c|c|c|c|c|c|c|c|c|c|c|c|c|}
 \hline \hline
 Data & \multicolumn{2}{|c}{Power Cost} & \multicolumn{2}{|c}{Wire Length w.
 LS} & \multicolumn{2}{|c}{LS Number}&\multicolumn{2}{|c}{ILO(\%)} & \multicolumn{2}{|c}{White Space(\%)} & \multicolumn{2}{|c|}{Run Time(s)}\\
 \cline{2-13}
 & \cite{GLSVLSI09YU} & MVLSAF & \cite{GLSVLSI09YU} & MVLSAF & \cite{GLSVLSI09YU} & MVLSAF & \cite{GLSVLSI09YU} & MVLSAF & \cite{GLSVLSI09YU}& MVLSAF & \cite{GLSVLSI09YU} & MVLSAF\\
 \hline
 n10  & 189142 & 162794 & 15504  & 16474  & 9   & 11 & 0.37 & 0.12 & 10.96 & 11.54 &  3.36 & 3.96 \\
 \hline
 n30  & 146483 & 138463 & 43265  & 45388  & 25  & 42 & 0.07 & 0.21 & 14.28 & 17.63 & 21.0  & 19.83\\
 \hline
 n50  & 143596 & 133564 & 94622  & 93296  & 114 & 151 & 0.28 & 0.50 & 22.63 & 22.95 & 41.37 & 49.35\\
 \hline
 n100 & 135607 & 120885 & 185382 & 181280 & 153 & 167 & 0.49 & 0.34 & 27.05 & 26.07 & 436.65& 414.7\\
 \hline
 n200 & 129615 & 117538 & 349562 & 344111 & 203 & 248 & 0.44 & 0.46 & 36.35 & 34.84 & 1980.4& 2036.4\\
 \hline
 n300 & 216554 & 206354 & 552616 & 568364 & 366 & 417 & 0.49 & 0.53 & 37.73 & 38.54 & 2384.3 & 2377.2 \\
 \hline
 Avg  & 160166 & 146599 & 206825 & 208152 & 145 & 172 & 0.36 & 0.36 & 24.83 & 25.26 & 811.2 & 816.8 \\
 \hline
 Diff &  -     & -8.5\% &  -   & +0.6\% &  -    & +18.6\% &  -     & $\pm$ 0\% &  -    & +1.7\% &  -     & +0.7\% \\
 \hline \hline
\end{tabular}
\end{table*}

\begin{table*}[tb]\label{table:result2}
 \centering
 \caption{Experimental Results with More Legal Working Voltage }
 \footnotesize
 \begin{tabular}{|c|c|c|c|c|c|c|c||c|c|c|c|c|c|c|c|}
   \hline \hline
   Data & $k$ & \tabincell{c}{Power\\Cost} & \tabincell{c}{Wire\\Length} & \tabincell{c}{LS\\Num} &
   \tabincell{c}{ILO\\(\%)} & \tabincell{c}{Dead\\Space(\%)} &
   \tabincell{c}{Run\\Time(s)} &
   Data & $k$ & \tabincell{c}{Power\\Cost} & \tabincell{c}{Wire\\Length} & \tabincell{c}{LS\\Num} &
   \tabincell{c}{ILO\\(\%)} & \tabincell{c}{Dead\\Space(\%)} &
   \tabincell{c}{Run\\Time(s)} \\
   \hline
   n10  & 2 & 189142 & 15341 & 9 & 0.10 & 10.21 & 3.05 &
   n100 & 2 & 133775 & 178685 & 122 & 0.17 & 26.45 & 431.7\\
        & 3 & 163352 & 16386  & 10  & 0.13 & 11.58 & 3.03 &
        & 3 & 131394 & 180023 & 150 & 0.50 & 26.8  & 438.05 \\
        & 4 & 162794 & 16474  & 11  & 0.12 & 11.54 & 3.96 &
        & 4 & 120885 & 181280 & 167 & 0.34 & 26.07 & 414.7 \\
   \hline
   n30  & 2 & 146483 & 42591 & 25 & 0.1 & 15.0 & 21.5 &
   n200 & 2 & 127044 & 344010 & 204 & 0.42 & 35.64 & 1955.8 \\
        & 3 & 139466 & 45103 & 42 & 0.32 & 15.85 & 20.82 &
        & 3 & 112801 & 331627 & 242 & 0.55 & 35.44 & 1949.4\\
        & 4 & 138463 & 45388 & 42   & 0.21 & 17.63 & 19.83 &
        & 4 & 117538 & 344111 & 248 & 0.46 & 34.84 & 2036.4 \\
   \hline
   n50  & 2 & 144489 & 93459  & 104 & 0.25 & 22.05 & 49.21 &
   n300 & 2 & 223574 & 548502 & 321 & 0.23 & 37.88 & 2363.9 \\
        & 3 & 132199 & 94105  & 130 & 0.37 & 22.72 & 51.10 &
        & 3 & 218636 & 556718 & 389 & 0.44 & 37.14 & 2390.2\\
        & 4 & 133564 & 93296  & 151 & 0.50 & 22.95 & 49.35 &
        & 4 & 206354 & 568364 & 417 & 0.53 & 38.54 & 2377.2\\
   \hline \hline
 \end{tabular}
\end{table*}
We implemented algorithm MVLSAF in the C++ programming language and
executed on a Linux machine with a 3.0GHz CPU and 1GB Memory.
Fig. \ref{result} shows the experimental results of the benchmarks
n50 and n100.

We use CBL\cite{CBL04} to represent every floorplan generated.
Besides, all the multi-pin nets are decomposed into a set of
source-sink two-pin nets. Cost function in floorplanning is: $\Phi =
\lambda_AA + \lambda_WW + \lambda_PP + \lambda_RR + \lambda_NN$,
where $A$ and $W$ represent the floorplan area and wire length; $P$
represents the total power consumption; $R$ represents the power
network resource; and $N$ records the number of level shifters that
can not be assigned.

The previous work \cite{GLSVLSI09YU} is the recent one in handling
floorplanning problem considering voltage  and level-shifter
assignment. We performed our algorithm MVLSAF and VLSAF in
\cite{GLSVLSI09YU} on the same test circuits, which are based on the
GSRC benchmarks adding power and delay specifications.
For each test circuit, we set $k$ as 4, and run MVLSAF and VLSAF 5
times. Table II lists the average results. The column Power Cost
means the actual power consumption. When allowing four legal working
voltages, MVLSAF can save 8.5\% power while not deteriorating
wirelength, dead space and run time. The column ILO and the column
LS Number show that using more accurate model in level shifter
assignment, even level shifters number increases 18.6\% ILO does not
increase.

In order to demonstrate the effectiveness of our approach, we have
done three  sets of experiments in which the number of legal working
voltage for each module is set two, three and four. The detailed
results are listed in Table III.

\begin{figure}[tb]
\centering
\includegraphics[width=0.23\textwidth]{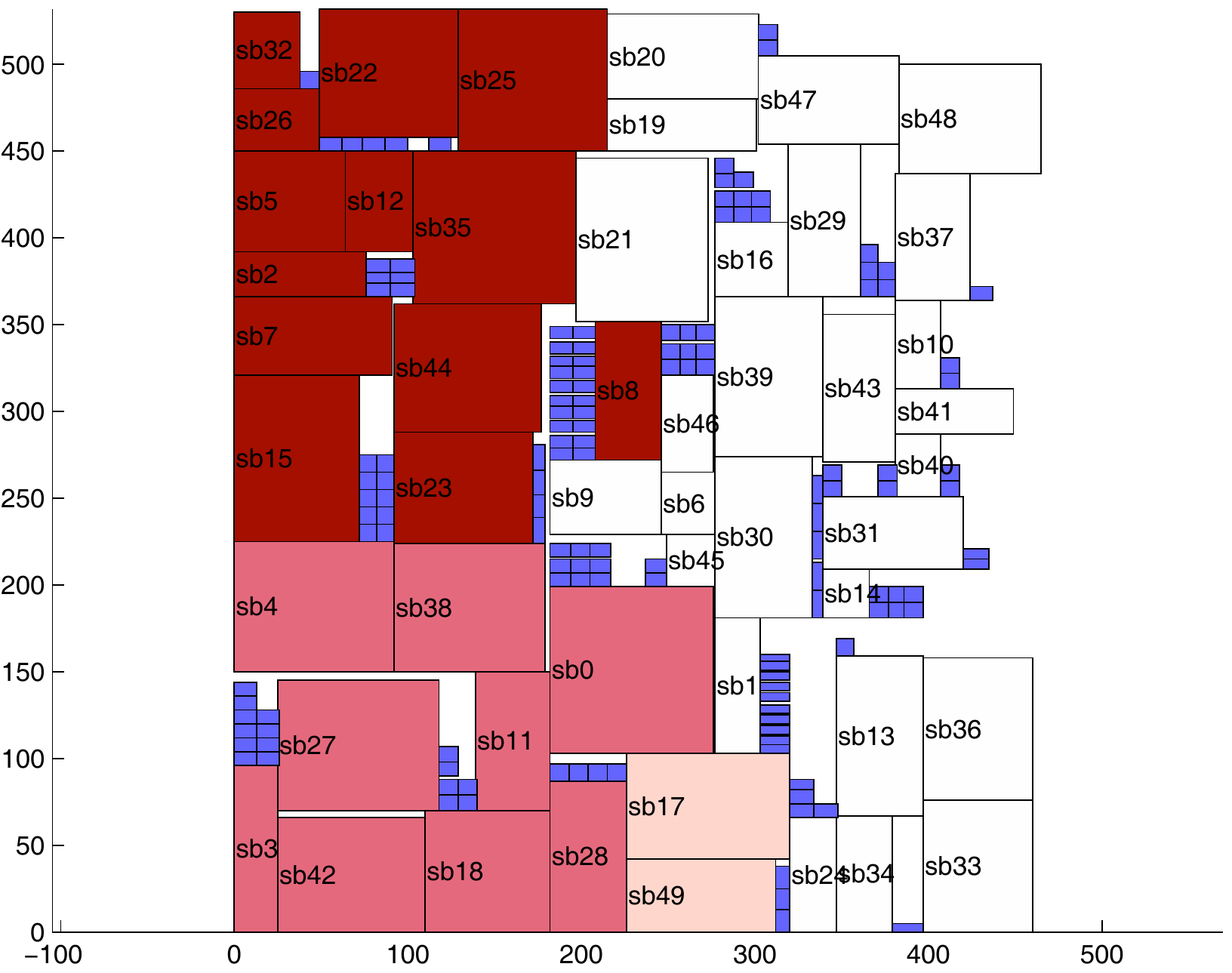}
\includegraphics[width=0.23\textwidth]{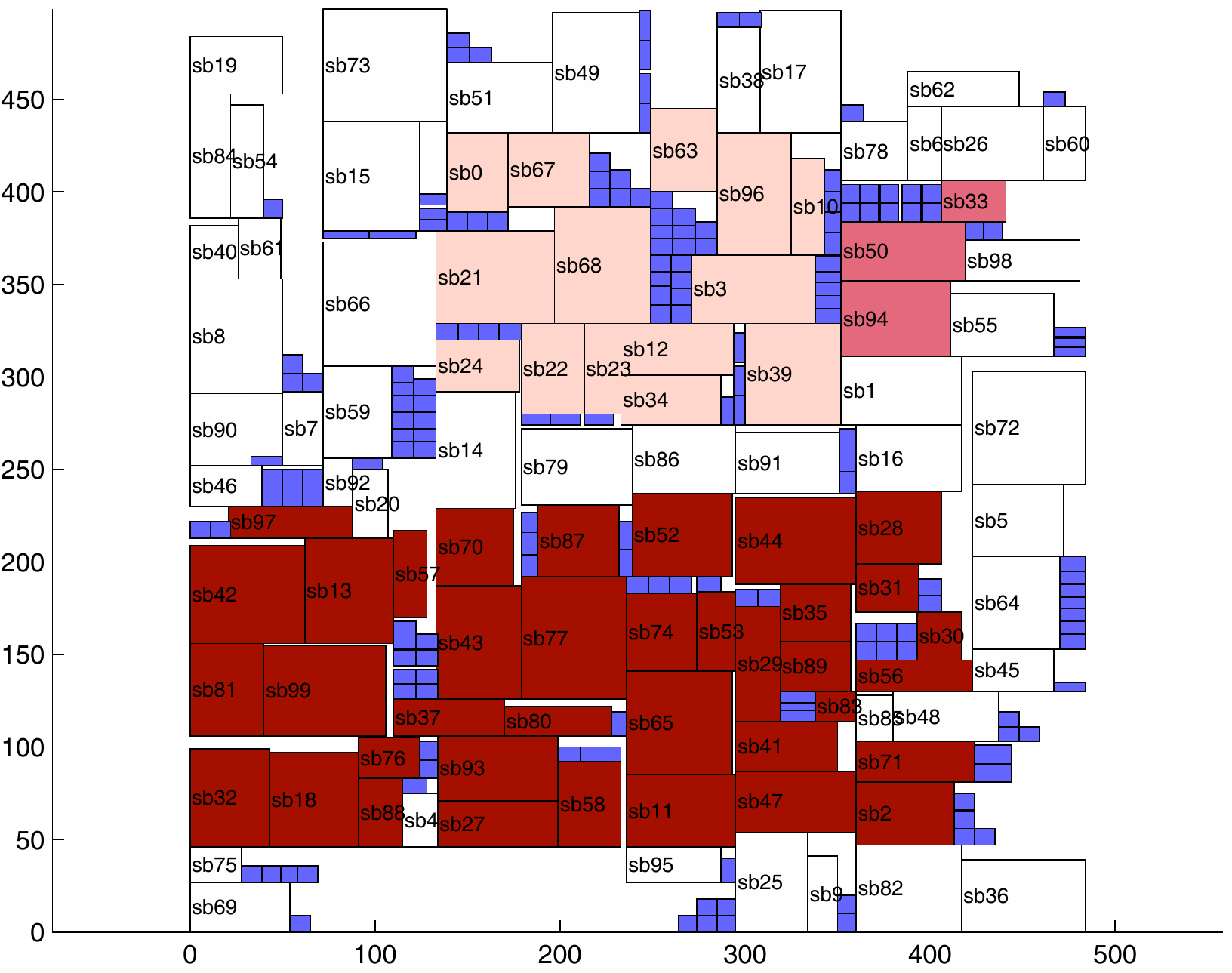}
\caption{Experimental results of  n50 and n100. Four legal working
voltages. Modules in the same voltage are nearly clustered together
to reduce the power-network resource, and level shifters (small dark
blocks) are inserted in white spaces.}\label{result}
\end{figure}

\section{\textbf{CONCLUSIONS}}

We have extended framework in \cite{GLSVLSI09YU} to solve
multi-voltage and level shifter assignment problem: a convex cost
network flow algorithm to assign arbitrary number of legal working
voltages; a minimum cost flow algorithm to handle level shifter
assignment. Experimental results have shown that our framework is
effective in reducing power cost while considering level shifters'
positions and areas.


\begin{thebibliography}{9}
\footnotesize

\bibitem{ICCAD02}
David~Lackey, Paul~Zuchowski and J.~Cohn.
\newblock Managing power and performance for system-on- chip designs using
  voltage islands.
\newblock {\em ICCAD}, pages 195--202, 2002.

\bibitem{CBL04}
Xianlong~Hong, Sheqin~Dong.
\newblock Non-slicing floorplan and placement using corner block list
  topological representation.
\newblock {\em IEEE Transaction on CAS}, 51:228--233, 2004.

\bibitem{FLOW03}
R.K. Ahuja, D.S. Hochbaum, and J.B. Orlin
\newblock Solving the convex cost integer dual network flow problem.
\newblock {\em Management Science}, 49(7):950-964, 2003

\bibitem{ICCD05}
W.L.Hung, G.M.Link and J.Conner.
\newblock Temperature-aware voltage islands architecting in system-on-chip
  design.
\newblock {\em ICCD}, 2005.

\bibitem{ICCAD06LEE}
W.P.Lee and Y.W.Chang.
\newblock Voltage island aware floorplanning for power and timing optimization.
\newblock {\em ICCAD}, pages 389--394, 2006.

\bibitem{ICCAD07MA}
Q.Ma and F.Y.Young.
\newblock Voltage Island-Driven Floorplanning.
\newblock {\em ICCAD}, pages 644--649, 2007.

\bibitem{DAC08}
D.Sengupta and R.Saleh.
\newblock Application-driven Floorplan-Aware Voltage Island Design.
\newblock {\em DAC}, pages 155--160, 2008.

\bibitem{ICCAD08MA}
Q.Ma and F.Y.Young.
\newblock Network flow-based power optimization under timing constraints in
  MVS-driven floorplanning.
\newblock {\em ICCAD}, 2008.

\bibitem{GLSVLSI09YU}
Bei Yu, Sheqin Dong, Satoshi GOTO and Song Chen.
\newblock Voltage-Island Driven Floorplanning Considering
Level-Shifter Positions.
\newblock {\em GLSVLSI}, 2009.

\bibitem{ASPDAC07}
W.K.Mak and J.W.Chen.
\newblock Voltage island generation under performance requirement for soc
  designs.
\newblock {\em ASP\_DAC}, 2007.

\bibitem{ICCAD07LEE}
W.P.Lee and Y.W.Chang.
\newblock An ILP algorithm for post-floorplanning voltage-island generation
  considering power-network planning.
\newblock {\em ICCAD}, pages 650--655, 2007.

\bibitem{ISPD09LEE}
W.P.Lee, D.Marculescu and Y.W.Chang.
\newblock Post-Floorplanning Power/Ground Ring Synthesis for
Multiple-Supply-Voltage Designs.
\newblock {\em ISPD}, pages 5--12, 2009.

\end{thebibliography}
\end{document}